\documentclass[letterpaper,10pt]{article} 
\usepackage{multicol}
\usepackage{opticameet3} %% use version 3 for proper copyright statement

\newcommand\authormark[1]{\textsuperscript{#1}}

\usepackage{amsmath,amssymb}
\usepackage[colorlinks=true,bookmarks=false,citecolor=blue,urlcolor=blue]{hyperref} %pdflatex

\begin{document}

\title{Timing Recovery for Point-to-Multi-Point Coherent Passive Optical Networks}

% \author{Author name(s)}
% \address{Author affiliation and full address}
% \email{e-mail address}
%%Uncomment the following line to override copyright year from the default current year.
%\copyrightyear{2022}

\author{Ji Zhou,\authormark{1,*} Jinyang Yang,\authormark{1}  Haide Wang,\authormark{1} Jianrui Zeng,\authormark{1} and Changyuan Yu\authormark{2}}

\address{\authormark{1} Department of Electronic Engineering, Jinan University, Guangzhou 510632, China\\
\authormark{2}The Hong Kong Polytechnic University Shenzhen Research Institute, Shenzhen 518057, China}

\email{\authormark{*}E-mail: zhouji@jnu.edu.cn} %% email address is required

\begin{abstract}
We propose a timing recovery for point-to-multi-point coherent passive optical networks. The results show that the proposed algorithm has low complexity and better robustness against the residual chromatic dispersion.
\end{abstract}

\section{Introduction}
Beyond 100Gb/s passive optical networks (PONs) will be required to satisfy the bandwidth demand in the near future. When the data rate is beyond 100Gb/s, intensity modulation and direct detection optical systems have limited receiver sensitivity and limited chromatic dispersion (CD) tolerance. Therefore, coherent optical technologies become a more attractive solution for the beyond 100Gb/s PON. Digital signal processing (DSP) is required to deal with channel distortions for recovering the signal in the coherent PON. However, if the CD compensation is used at optical network units (ONUs) \cite{CD compensation}, the complexity is unacceptable. Therefore, the CD can be pre-compensated at the optical line terminal (OLT) to relieve stress on ONUs.

Due to the point-to-multi-point (P2MP) structure of PON, the downstream signal is broadcast from the OLT to all ONUs, which can always receive the signal to synchronize the clock at the ONUs. The timing recovery is a crucial part of the DSP, which can be used to correct the sampling error. However, the residual CD after the pre-compensation at the OLT will lead to an unstable performance of the traditional timing recovery \cite{Gardner, Godard, Fourth}. In this paper, we propose a timing recovery integrating with simplified CD compensation for the P2MP coherent PON. The results show that the coherent time-and-frequency division multiple access (TFDMA)-PON with a lower baud-rate subcarrier is more robust against CD. For the coherent TFDMA-PON with a high baud-rate subcarrier, the proposed timing recovery algorithm has low complexity and better robustness against the residual CD.

\begin{figure}[!b]
\centering
\includegraphics[width=\linewidth]{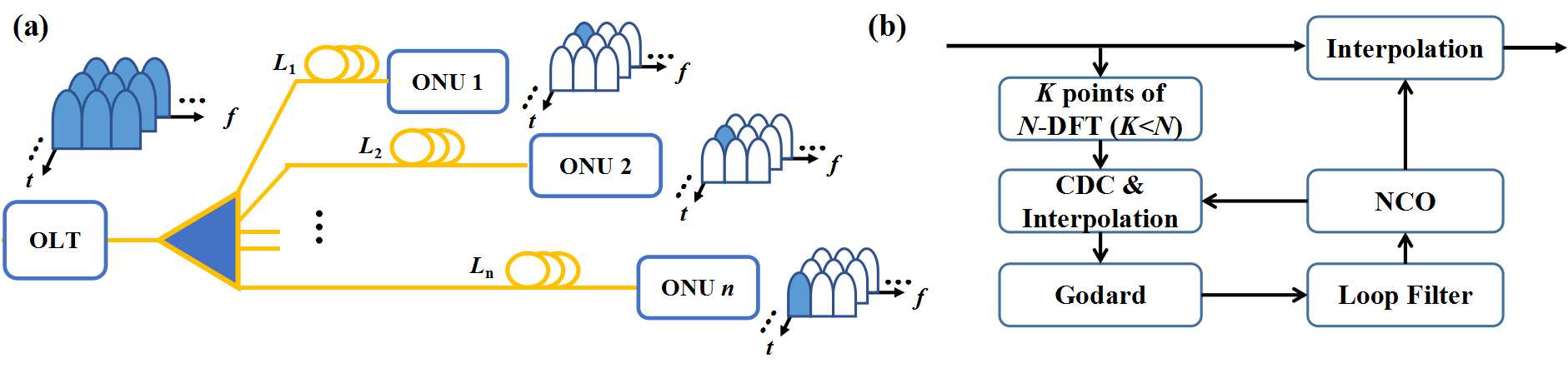}
\caption{(a) Schematic diagram of P2MP PON in downstream transmission. (b) Structure of the proposed timing recovery algorithm, only a part of the frequency components are required to compensate CD and sent to timing recovery to estimate the timing error.}
\label{fig_1}
\end{figure}

\section{Timing Recovery for P2MP Coherent PON}

The schematic diagram of P2MP PON using digital subcarrier multiplexing (DSCM) in the downstream transmission is shown in Fig. \ref{fig_1}(a), which can provide TFDMA \cite{H. Wang}. Since the subcarriers are divided into time slots and allocated to ONUs with different transmission distances, the accumulated CD for pre-compensation is different in different time slots. If the received signal is at the time slots allocated to other ONUs, the residual CD will cause an unstable performance of timing recovery. Therefore, the residual CD will make it unable to track the sampling phase accurately at the time slots of other ONUs. It will lead to a long convergence time of timing recovery in their own time slots, which may result in block error and the poor performance of subsequent algorithms. 

The structure of the proposed simplified CD-compensated timing recovery is shown in Fig. \ref{fig_1}(b). According to the principle of the Godard algorithm, the clock information only exists at frequency components around the half-baud rate. It illustrates that only a part of the components provides useful contributions to the timing estimation. The other components are zero in magnitude and only disturb the estimation with additional noise. Neglecting the useless components not only reduces the computational complexity but also improves the performance of the timing recovery. Therefore, there is no need to compensate for the residual CD for all frequency components of the data. The residual CD of a few frequency components around the half-baud rate are compensated in the frequency domain \cite{Kudo R}. Then the CD-compensated components are sent to the feedback loop of timing recovery to calculate the timing error. Finally, the estimated timing error is used to recover the timing phase for the data by interpolation.

\section{Results and Discussions}
\begin{figure}[!t]
\centering
\includegraphics[width=\linewidth]{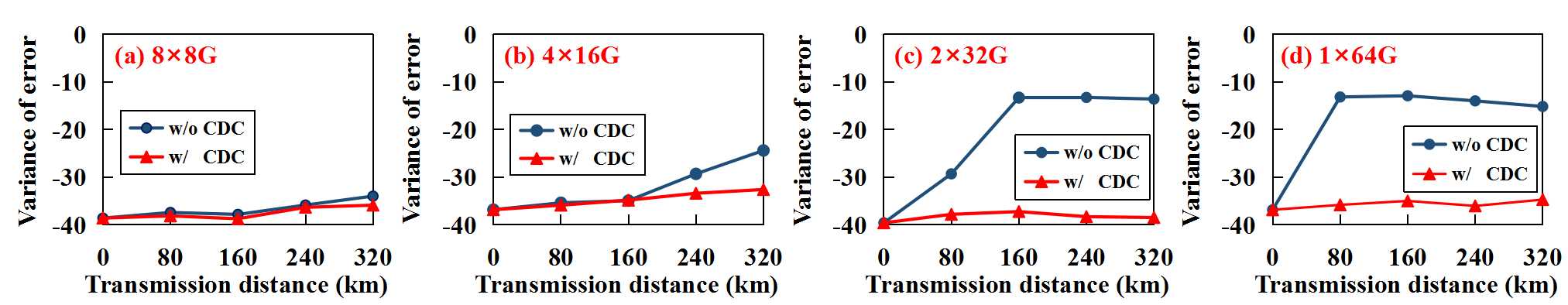}
\caption{Variance of timing estimation error versus transmission distance for (a) $8$Gbaud/SC$\times8$SCs, (b) $16$Gbaud/SC$\times4$SCs, (c) $32$Gbaud/SC$\times2$SCs DSCM signal and (d) $64$Gbaud 16QAM signal, respectively.}
\label{fig_2}
\end{figure}

The influence of CD on timing recovery is evaluated for the DSCM signals using $16$ quadrature amplitude modulation (QAM) with $8$-Gbaud/SC$\times8$-SCs, $16$-Gbaud/SC$\times4$-SCs, $32$-Gbaud/SC$\times2$-SCs and $64$-Gbaud 16QAM signal. The CD is set to 16ps/nm/km. The received signal is oversampled at two samples per symbol. The DFT size $N$ is 128 and the roll factor is 0.1. Thus, only $K = 12$ points are needed to be transformed to the frequency domain to compensate for CD and estimate the timing phase, which can reduce computational complexity effectively. Fig. \ref{fig_2} shows the variance of the timing estimation error versus transmission distance. For the $16$-Gbaud subcarrier, the variance of timing error becomes slightly larger. If the baud rate is $32$ Gbaud or $64$ Gbaud, the variance of timing error increases sharply. As a result, the signal with a higher baud rate is more susceptible to the influence of CD on timing recovery without the low-complexity CD compensation in the timing recovery loop. 

\begin{figure}[htbp]
\centering
\includegraphics[width=\linewidth]{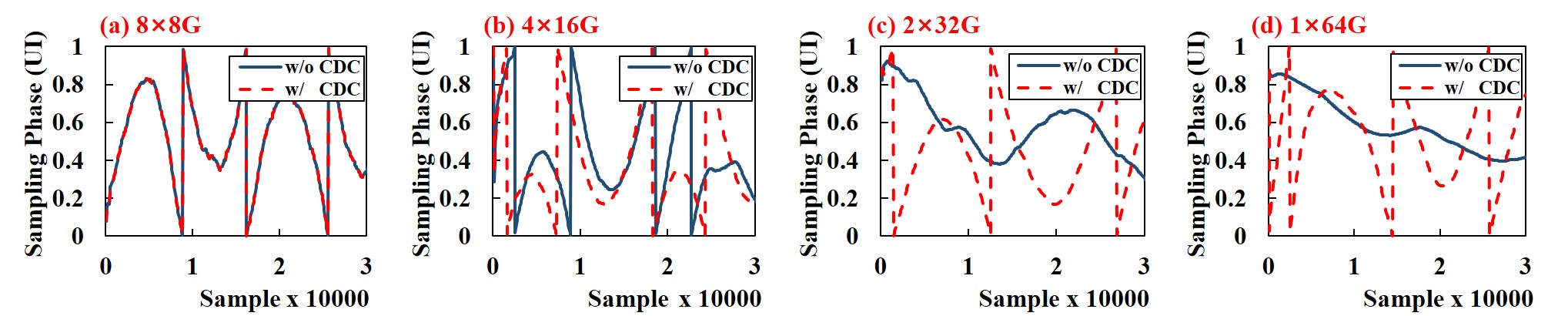}
\caption{Estimated sampling phase of (a) $8$Gbaud/SC$\times8$SCs, (b) $16$Gbaud/SC$\times4$SCs, (c) $32$Gbaud/SC$\times2$SCs DSCM signal and (d) $64$Gbaud 16QAM signals with $320$km $\times$ 16ps/nm/km accumulated CD, respectively.}
\label{fig_3}
\end{figure}

A sampling phase error is introduced into the system over $320$-km transmission. Fig. \ref{fig_3} shows the estimated timing phase error of the proposed algorithm without and with the low-complexity CD compensation in the timing recovery loop. When the baud rate is $8$ Gbaud, there is almost no impact on the timing recovery. For the $16$-Gbaud subcarrier,  Godard timing recovery without CD compensation cannot estimate the timing phase accurately. If the baud rate is $32$ Gbaud or $64$ Gbaud, Godard timing recovery without CD compensation has lost the ability to track the timing phase sustainably. In conclusion, the coherent TFDMA-PON with a lower baud-rate subcarrier is more robust against CD. For the coherent TFDMA-PON with a high baud-rate subcarrier, the proposed timing recovery algorithm can track the timing phase error accurately in the presence of a large accumulated CD.

\section{Conclusions}
In this paper, we analyze the effect of the residual CD on the timing recovery and propose a timing recovery for P2MP coherent PONs. The results show that the coherent PON with a lower baud-rate subcarrier for TFDMA is more robust against CD. For the coherent TFDMA-PON with a high baud-rate subcarrier, the proposed timing recovery algorithm has low complexity and better robustness against the residual CD.

\emph{This work is supported in part by the National Natural Science Foundation of China (62005102); Key Basic Research Scheme of Shenzhen Natural Science Foundation (JCYJ20200109142010888); Hong Kong Scholars Program (XJ2021018).}

\end{document}